\documentclass[12pt,aps,prapplied,reprint, showpacs,showkeys, superscriptaddress]{revtex4-1}
\usepackage{newcent}

\usepackage[T1]{fontenc}
\usepackage[latin9]{inputenc}
\usepackage{textcomp}
\usepackage{amsmath}
\usepackage{commath}
\usepackage{csquotes}
\usepackage{graphicx, subfigure}
\usepackage{amssymb}
\usepackage{amsbsy}
\usepackage{textcomp}
\usepackage{eucal}
\usepackage{siunitx}
\usepackage{textcomp}
\usepackage{pslatex}
\usepackage{relsize}
\usepackage{bm}
\usepackage{color}
\usepackage{hyperref}
\hypersetup{colorlinks=true,linkcolor=blue,citecolor=blue,urlcolor=blue}
\usepackage[section]{placeins} 

\usepackage[normalem]{ulem}

\begin{document}
\title{Nanoparticle size threshold for magnetic agglomeration and associated hyperthermia performance}
\author{D. Serantes}
\email[E-mail: ]{david.serantes@usc.gal}
\affiliation{Applied Physics Department and Instituto de Investigaci\'ons Tecnol\'oxicas, Universidade de Santiago de Compostela, 15782 Santiago de Compostela, Spain}
\author{D. Baldomir}
\affiliation{Applied Physics Department and Instituto de Investigaci\'ons Tecnol\'oxicas, Universidade de Santiago de Compostela, 15782 Santiago de Compostela, Spain}

\begin{abstract}
The likelihood of magnetic nanoparticles to agglomerate is usually estimated through the ratio between magnetic dipole-dipole and thermal energies, thus neglecting the fact that, depending on the magnitude of the magnetic anisotropy constant ($K$), the particle moment may fluctuate internally and thus undermine the agglomeration process. Based on the comparison between the involved timescales, we study in this work how the threshold size for magnetic agglomeration ($d_{aggl}$) varies depending on the $K$ value. Our results suggest that small variations in $K$ -due to e.g. shape contribution-, might shift $d_{aggl}$ by a few nm. A comparison with the usual \textit{superparamagnetism} estimation is provided, as well as with the energy competition approach. In addition, based on the key role of the anisotropy in the hyperthermia performance, we also analyse the associated heating capability, as non-agglomerated particles would be of high interest for the application. 

\end{abstract}

\maketitle

\section{Introduction}

Based on the possibility to achieve local actuation by a harmless remote magnetic field, magnetic nanoparticles are very attractive candidates for novel medical  applications \cite{wu2019,Colombo2012}. Particularly iron oxides, based on their good biocompatibility \cite{Ling2013}, have been the subject of intense research in recent years, for example for magnetic hyperthermia cancer therapy \cite{SOETAERT2020,ABENOJAR2016} or drug release \cite{FortesBrollo2020,Thorat2017}.

A key aspect to consider when dealing with magnetic nanoparticles for biomedical applications is the agglomeration likelihood, as it could affect not only the metabolising process but also the magnetic properties by changing the interparticle interactions \cite{ROJAS2017}. Considering for example magnetic hyperthermia, it is known that the particles tend to agglomerate when internalized by the cells and that such may lead to a decrease of the heating performance \cite{Mejias2019}. However, the opposite behaviour has also been reported, with an increase of the heat release if the particles form chains \cite{Serantes2014}. In general, accounting for the effect of interparticle dipolar interactions is of primary importance for a successful application \cite{Gutierrez2019}.

The complex role of the interparticle interactions often prompts researchers to the use of \textit{superparamagnetic} (SPM) particles, with the idea that the rapid internal fluctuation of the particles' magnetic moments shall prevent their agglomeration. Thus, in first approximation one could be tempted to consider that agglomeration will not occur for particles with blocking temperature ($T_B$) below the desired working temperature, since for $T>T_{B}$ the particles are in the SPM state (i.e. they behave paramagnetic-like). However, it must be kept in mind that behaving SPM-like is not an absolute term, but it is defined by the experimental timescale. Thus, regarding agglomeration, a particle could be referred to as SPM if its N\'eel relaxation time, $\tau_{N}$, is smaller than the characteristic timescales that allow agglomeration, i.e. diffusion ($\tau_{diff}$) and rotation ($\tau_{B}$)  \cite{Balakrishnan2020}. These are given by \begin{equation}\label{Neel}
    \tau_N=\frac{\sqrt{\pi}}{2}{\tau_{0}}\frac{e^{\frac{KV}{k_{B}T}}}{(\frac{KV}{k_{B}T})^{\frac{1}{2}}},
\end{equation} \begin{equation}\label{diffusion}
    \tau_{diff}=\frac{x^26\pi{\eta}R_{hyd}}{k_{B}T},
\end{equation} and 
\begin{equation}\label{Brown}
    \tau_{B}=\frac{3{\eta}V_{hyd}}{k_{B}T},
\end{equation} respectively, where $\tau_0=10^{-9}$ s, $K$ is the uniaxial anisotropy constant and $V$ the particle volume; $k_{B}$ is the Boltzmann constant, $x$ the particle diffusion distance, and $\eta$ the viscosity of the embedding media; $R_{hyd}$ and $V_{hyd}$ are the hydrodynamic radius and volume, respectively, defined by the particle size plus a nonmagnetic coating of thickness $t_{nm}$. For simplicity we consider spherical particles of diameter $d$.

The objective of this work is to estimate the size threshold for magnetic agglomeration, $d_{aggl}$ (i.e. size for which $\tau_{N}>\tau_{diff},\tau_{B}$, so that agglomeration is likely) in terms of $K$. Focusing on magnetite-like parameters based on its primary importance for bioapplications, we will consider different \textit{effective} $K$ values, which can be ascribed to dominance of shape anisotropy over the magnetocrystalline one \cite{Usov2010,Vallejo-Fernandez2013}. Comparison will be made with the usual estimate of agglomeration likelihood: the ratio between the dipolar energy of parallel-aligned moments and thermal energy \cite{Andreu2011,SATOH1996},
\begin{equation}\label{gamma}
    \Gamma=\frac{\mu_{0}(M_{S}V)^2}{2{\pi}{l_{cc}}^3k_{B}T},
\end{equation} in the limit case of touching particles (i.e. $l_{cc}=d$). In Eq. \eqref{gamma}, $\mu_{0}=1.256*10^{-6} Tm/A$ is the permeability of free space, $M_{S}$ the saturation magnetization, and $l_{cc}$ the center to center interparticle distance. Note that eq. \eqref{gamma} does not consider $K$, despite its key role in governing the magnetization behaviour. Then, the hyperthermia properties for the obtained $d_{aggl}$ will be studied. It must be recalled here the double role of $K$ in the heating performance, as it determines both the maximum achievable heating \cite{Dennis2015, Conde-Leboran2015} and the effectiveness in terms of field amplitude \cite{Munoz-Menendez2017}; for completeness, this double role of $K$ will also be briefly summarized. Please note that we are using "agglomeration" referring to a \textit{reversible} process, distinct from the \textit{irreversible} "aggregation" \cite{Gutierrez2015}.

\section{Results and discussion}\label{results}
\subsection{Size threshold for magnetic agglomeration, $d_{aggl}$}\label{size}
\begin{figure}[h!]
\centering
\includegraphics[width = 1.0\columnwidth]{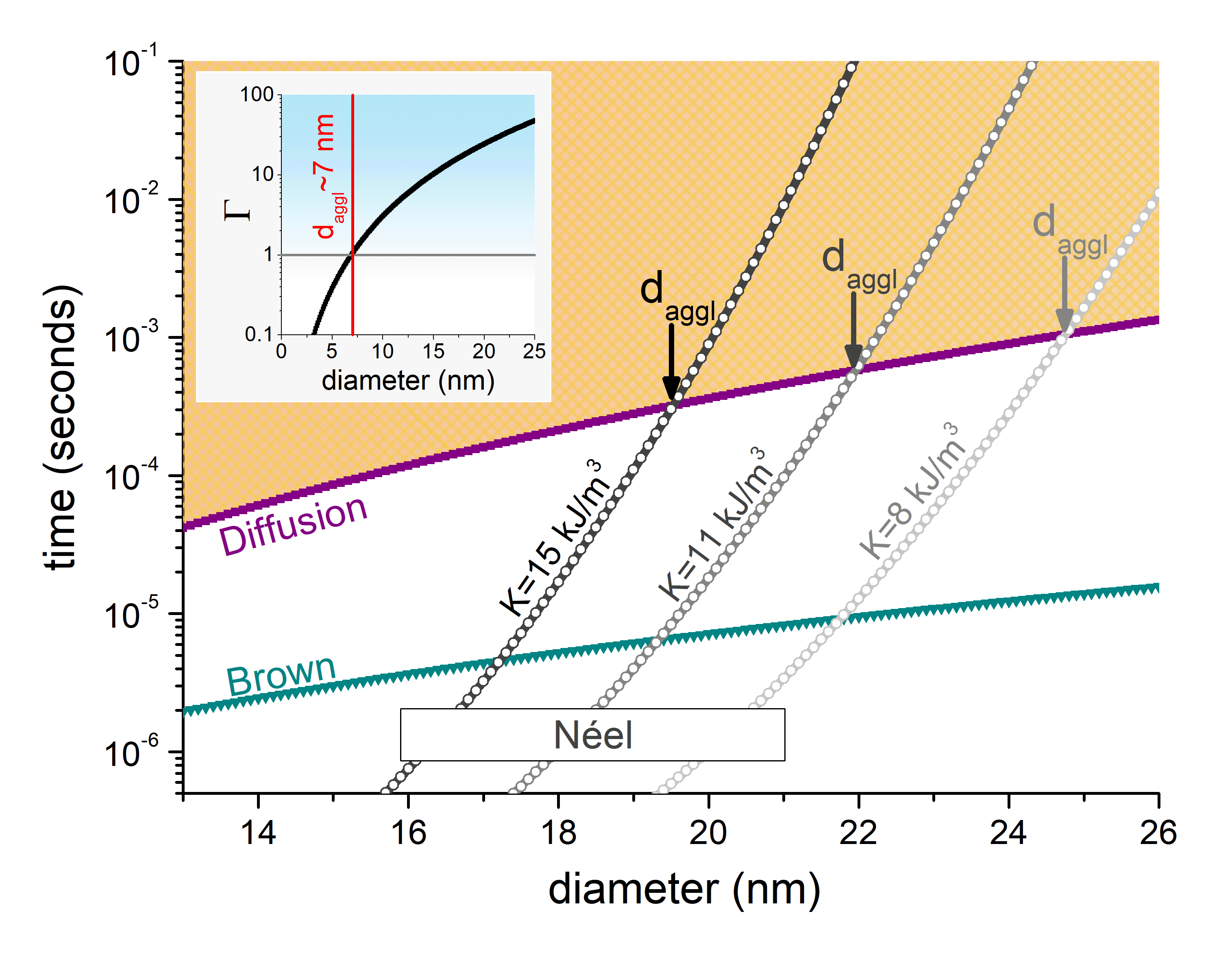}
\caption{Diffusion ($\tau_{diff}$; purple line), Brown ($\tau_B$; green line), and N\'eel relaxation time ($\tau_N$; grey lines), as a function of the particle diameter. The distinct $\tau_N$ curves correspond to the different $K$ values indicated. The dashed light-orange area indicates the range where agglomeration can be expected. The inset shows the size dependence of the $\Gamma$, which predicts agglomeration for sizes $d>~7$ nm.}
\label{fig:times}
\end{figure}

To estimate $d_{aggl}$ we followed the same approach as we did in Ref. \cite{Balakrishnan2020}: to compare the characteristic N\'eel, diffusion, and rotation times, to obtain $d_{aggl}$ as the size for which $\tau_{N}>\tau_{diff},\tau_{B}$. In eqs. \eqref{diffusion}-\eqref{Brown} we have at first set $t_{nm}=0$, and used $\eta = 0.00235$ kg/m*s, as in Ref. \cite{ROSENSWEIG2002}, which is comparable to that of HeLa cells for nm-scale dimensions \cite{Kalwarczyk2011}. We considered three cases for eq. \eqref{Neel}: $K=8, 11$, and $15$ $kJ/m^3$, i.e. values of the order found in the literature for magnetite particles \cite{Nguyen2020,Balakrishnan2020,Niculaes2017}. The diffusion distance in eq. \eqref{diffusion} is set as the interparticle distance at which the magnetostatic energy dominates over the thermal one, i.e. ${\Gamma>1}$ \cite{Balakrishnan2020}, so that:
\begin{equation}\label{distance}
    x=\left(\frac{\mu_{0}(M_{S}V)^2}{2{\pi}{k_{B}T}}\right)^{\frac{1}{3}}.
\end{equation} Note that while we have chosen $\Gamma=1$ to have a well defined criterion, agglomeration usually requires higher $\Gamma$ values \cite{Santiago-Quinones2013}. That is to say, we are searching for the lower $d_{aggl}$ boundary. With the same spirit, in eq. \eqref{gamma} we used $M_{S}=4.8*10^5$ $A/m$, i.e. the upper value for magnetite so that the interaction is, most likely, overestimated. The relaxation times as a function of the particle size are shown in Figure \ref{fig:times}.

In Figure \ref{fig:times} it is clearly observed how increasing $K$ leads to more stable moments, thus favouring agglomeration at smaller sizes (from $d_{aggl}\sim{25}$ nm for $K=8$ $kJ/m^3$, to $d_{aggl}\sim{20}$ nm for $K=15$ $kJ/m^3$). The inset shows the size dependence of $\Gamma$, which i) does not distinguish among particle characteristics (in terms of $K$, as previously mentioned), and; ii) predicts dominance of the dipolar energy for much smaller particle sizes, with $d_{aggl}\sim{7}$ nm. It is worth noting that the threshold value obtained for the $K=11$ $kJ/m^3$ case, $d_{aggl}\approx{22}$, is slightly bigger than the one previously reported in Ref. \cite{Balakrishnan2020}, for which $d_{aggl}\approx{21}$ nm. This is due to the larger $M_{S}$ value used here, which enhances the diffusion time (through the diffusion distance, eq. \eqref{distance}). Nevertheless, the great similarity despite the different $M_{S}$ values emphasizes the key role of the anisotropy in the agglomeration likelihood. The fact that so far we are not considering a nonmagnetic coating has a minor effect, as discussed next.

\begin{figure}[h!]
\centering
\includegraphics[width = 1.0\columnwidth]{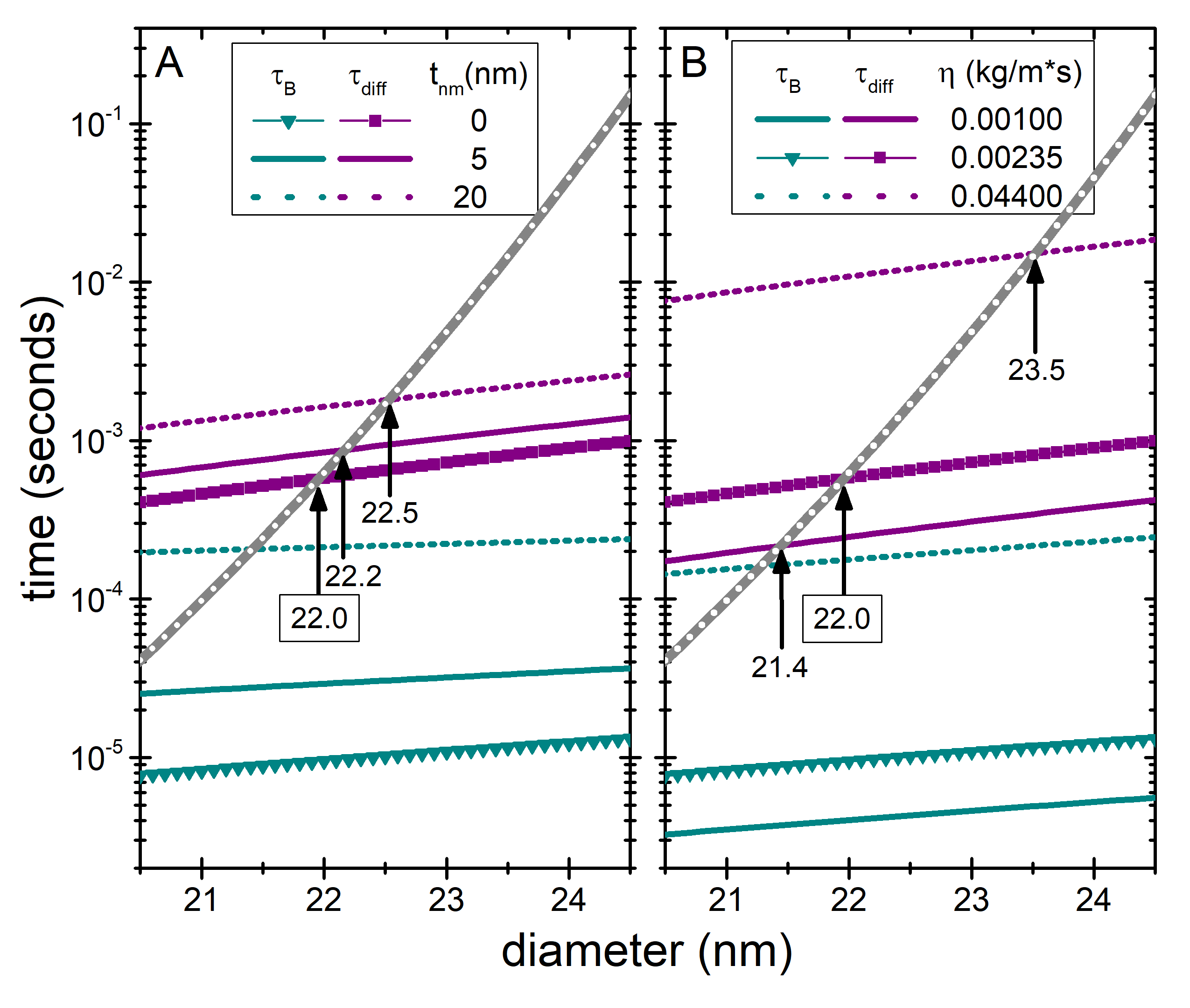}
\caption{Diffusion ($\tau_{diff}$; purple line), Brown ($\tau_B$; green line), and N\'eel relaxation time ($\tau_N$; grey line), as a function of the particle diameter, as in Figure \ref{fig:times}, but considering different thickness of the nonmagnetic coating (left A panal), or viscosity of the medium (right B panel). For simplicity, the results are focused on the $K=11$ $kJ/m^3$ and the original curves from Figure \ref{fig:times} are reproduced for guidance. The variations of $t_{nm}$ and $\eta$ are shown with solid and dotted lines, for the values displayed within each panel. The arrows and attached numbers indicate $t_{aggl}$, with the reference one (22.0 nm) highlighted.}
\label{fig:hydSIZEvisc}
\end{figure}

While we considered $t_{nm}=0$ in order to determine the boundary where clustering might appear, biomedical applications will always require a biocompatible nonmagnetic coating and therefore it is important to consider its role. That being said, the analysis shows that including a non-magnetic coating does not significantly modify the obtained threshold values: if considering $t_{nm}=5$ nm, $d_{aggl}$ increases just by $\sim{0.2}$ nm; and by $\sim{0.5}$ nm if $t_{nm}=20$ nm. This is illustrated in Figure \ref{fig:hydSIZEvisc}A.

A slightly larger influence is that of the viscosity of the embedding media, as illustrated in Figure \ref{fig:hydSIZEvisc}B. Considering for example that of water, $\eta = 0.001$ kg/m*s, it is observed a $0.6$ nm decrease from the average size. This value of viscosity is very significant because of being very similar to that of the cells cytoplasm, although it must be kept in mind that large variations can be observed within the same cell type and among different types of cells \cite{Wang2019}. A much higher viscosity would have a more significant effect, as illustrated for example with the macroscopic value of HeLa cells, $\eta = 0.044$ kg/m*s; nevertheless this values would be unrealistically high for the current particles, as such large $\eta$ would correspond to much bigger sizes (over $\sim{86}$ nm for HeLa cells) because of the size-dependent viscosity at the microscale \cite{Kalwarczyk2011}.

It is important to note that for the anisotropy values considered here, in all  cases the size threshold $d_{aggl}$ is always defined by the competition between diffusion and N\'eel times, as $\tau_{B}<\tau_{diff}$ for all cases shown in Figure \ref{fig:hydSIZEvisc}.

Next we will compare the predictions from the relaxation times with those obtained from \textit{zero field cooling/field cooling} (ZFC/FC) measurements, the common way to estimate SPM behaviour (and thus likely non-agglomeration). Thus, if associating the onset of SPM behaviour to the blocking temperature, estimated as $T_{B}=KV/25k_B$ \cite{Livesey2018}, the corresponding threshold size, $d_{T_{B}}$, is readily obtained. The comparison between the agglomeration thresholds predicted by both approaches at room temperature (i.e. setting $T_{B}=300$ $K$) is summarized in Table \ref{table-d_aggl}.

\begin{table}[!ht]
\caption{Agglomeration size thresholds obtained through the relaxation times approach ($d_{aggl}$) and through the ZFC/FC one ($d_{T_{B}}$), at room temperature for the three anisotropy cases of Figure \ref{fig:times}.}
\begin{center}
\begin{tabular}{c c c c c  c}
\hline
\hline
$K (kJ/m^3)$ & & $d_{aggl} (nm)$ & & $d_{T_{B}} (nm)$ \\
\hline
8 & & 24.8 & & 29.2 \\
11 & & 22.0 & & 26.2 \\
15 & & 19.5 & & 23.6 \\
\hline
\hline
\end{tabular}
\end{center} 
\label{table-d_aggl}
\end{table}

Table \ref{table-d_aggl} shows that, on average, the ZFC/FC approach predicts agglomeration to occur for sizes $\sim{4.2}$ nm bigger than the ones predicted by the relaxation times approach. In fact, the obtained $d_{T_B}$ values correspond to a lower boundary, as they were estimated considering the limit case of no applied field, which is not possible in real ZFC/FC experiments. In general, applying the field during the measurements will result in lower $T_B$ \cite{Goya2004,Nunes2005,BALAEV2017}, which would correspond to larger $d_{T_B}$ (at least for the monodisperse case considered here; polydispersity might result in more complex scenarios \cite{Chantrell2000,Kachkachi2000,Usov2011}).

\subsection{Associated heating performance}

\begin{figure}[h!]
\centering
\includegraphics[width = 1.0\columnwidth]{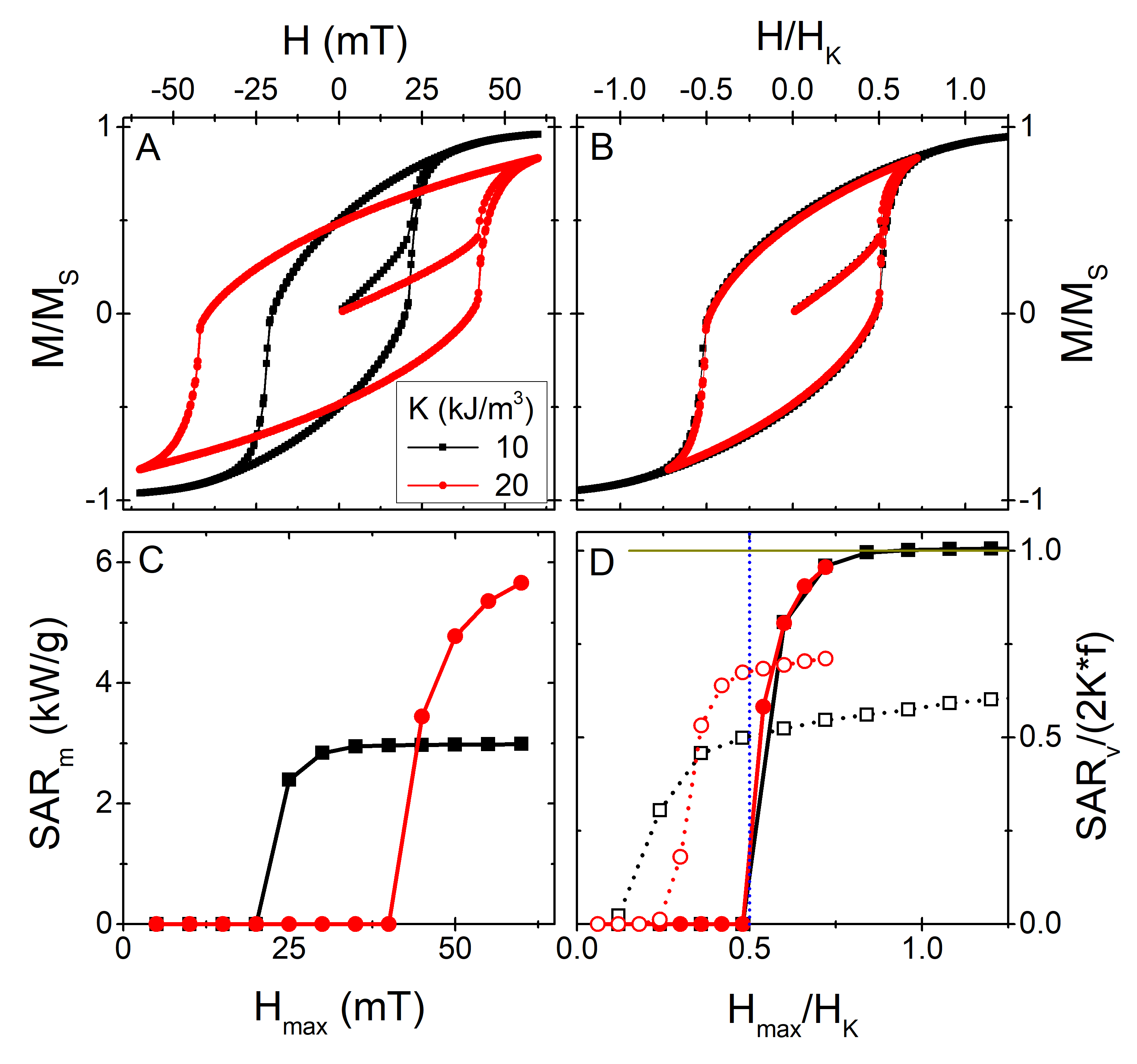}
\caption{A: Illustrative $M$ \textit{vs.} $H$ hysteresis loops of two systems of particles of same size ($d=20$ nm) and $M_S=480$ kA/m, but different $K$ (10 and 20 $kJ/m^3$, respectively), at $T=0$ K and for $H_{max}=25$ mT. B: Same data as in A, replotted in terms of $H/H_K$. C: SAR vs. $H_{max}$ for the two different particles, for $f=765$ kHz, at $T=0$ K. D: Same data as in panel C, replotted in terms of $SAR/2K*f$ and $H/H_K$; the curves with open symbols correspond to the $T=300$ K case. The vertical blue dotted line stands for the $\sim{0.5H_K}$ threshold of the random distribution \cite{Serantes2010}, and the horizontal solid dark-yellow line indicates the normalized maximum $SAR/(2K*f)$=1 limit case.}
\label{fig:scaling}
\end{figure}

Similar to its importance on the agglomeration likelihood, the anisotropy plays a principal role in defining the hyperthermia performance. On the one hand, it defines the maximum energy that can be dissipated \cite{Serantes2010,SOETAERT2020}: it is easy to see that for aligned easy axes the maximum hysteresis losses \textit{per loop} are $8K$ \cite{area8K} ($2K$ for the random easy axes distribution \cite{Conde-Leboran2015}). On the other hand, it settles the response to the applied field (of amplitude $H_{max}$) through the anisotropy field, defined as $H_{K}=2K/\mu_{0}M_{S}$ \cite{Serantes2010,Munoz-Menendez2017}. This double key-role is illustrated in Figure \ref{fig:scaling}, where the heating performance is reported in terms of the usual \textit{Specific Absorption Rate} parameter, SAR, as $SAR=A*f$, where $A$ stands for the area of the loop (hysteresis losses), and $f$ is the frequency of the AC field. The simulations were performed in the same way as in Ref. \cite{Balakrishnan2020}: we considered a random dispersion of monodisperse non-interacting nanoparticles (with the easy axes directions also randomly distributed), and simulated their response under a time varying magnetic field by using the standard Landau-Lifshitz-Gilbert equation of motion within the OOMMF software package \cite{oommf}; for the random thermal noise (to account for finite temperature) we used the extension module \textit{thetaevolve} \cite{oommf-thetaevolve}.

The results displayed in Figure \ref{fig:scaling} show how, same as the apparently different hysteresis loops (A panel) are scaled by the anisotropy field (B panel), the apparently different SAR \textit{vs.} $H_{max}$ trends scale if plotting $SAR/(2K*f)$ vs. $H_{max}/H_K$ (the $2f$ factor is just for normalisation). Note, however, that those results correspond to the Stoner-Wohlfarth-like case at $T=0$ K \cite{Lacroix2009}. In real systems with finite temperature, $K$ also defines -as previously discussed- the stability of the magnetization within the particle. Thus, the ideal $T=0$ K situation may vary significantly due to the effect of thermal fluctuations, as shown by the open symbols in Figure \ref{fig:scaling}D, which correspond to the $T=300$ K case for the two particle types considered. It is clearly observed how the strict $H_{max}\sim{0.5H_K}$ threshold does not hold, and that the SAR is much smaller than the maximum possible.

The results shown in Figure \ref{fig:scaling} illustrate well the the double role of the anisotropy on the heating performance. What is more, it must be kept in mind that the magnetic anisotropy is the only reason why small particles, such as the ones considered here of typical hyperthermia experiments (well described by the \textit{macrospin} approximation) release heat under the AC field: \textit{if no anisotropy were to exist, there would be no heating} (at least not for the frequencies and fields considered). This applies both to N\'eel and Brown heating, as with no anisotropy the magnetization would not transfer torque to the particle for its physical reorientation. Of course, larger sizes could display different heating mechanisms (due to non-coherent magnetization behaviour \cite{Usov2018} or even eddy currents \cite{Morales2020}), but that is not the present case.

\begin{figure}[h!]
\centering
\includegraphics[width = 1.0\columnwidth]{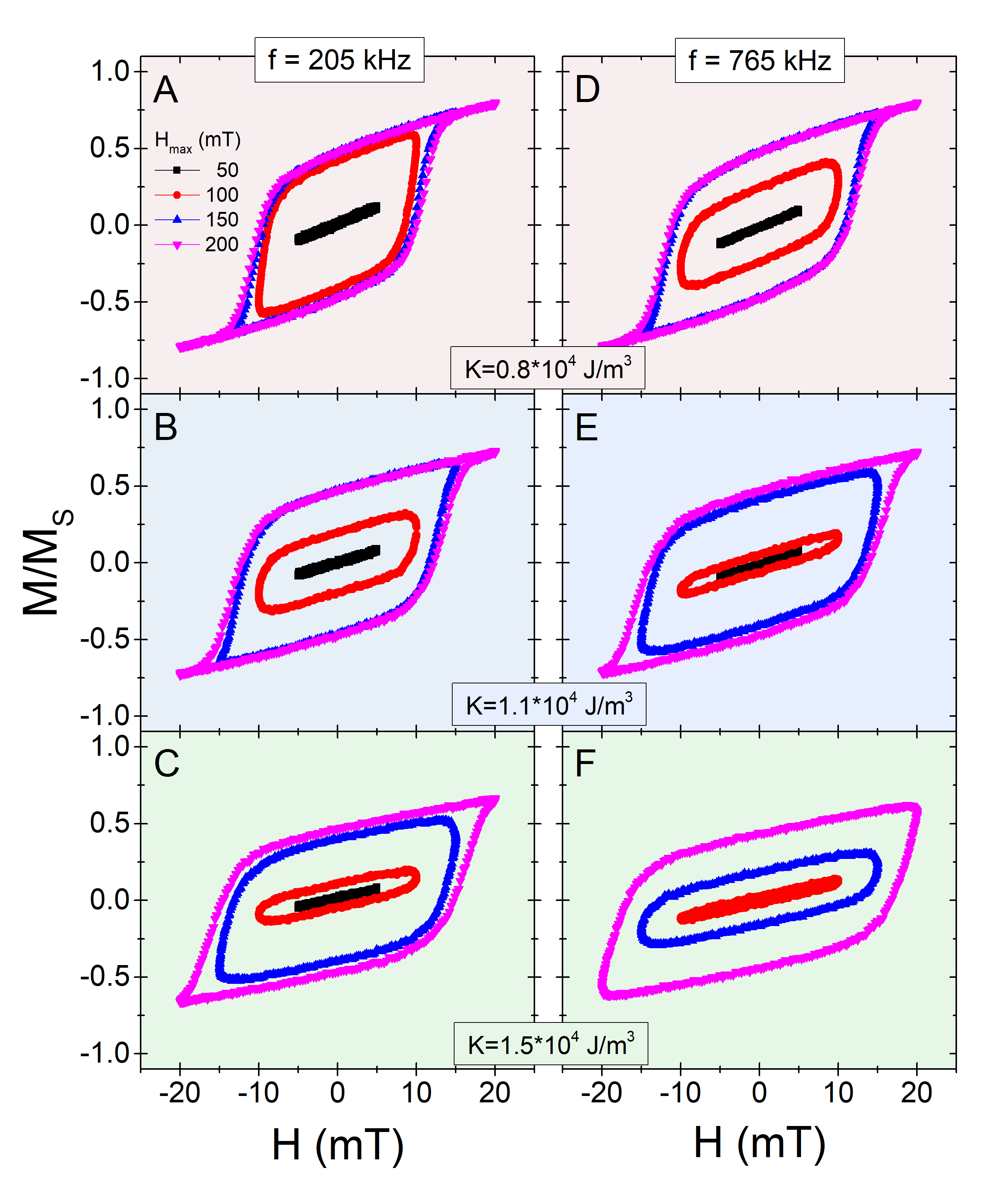}
\caption{$M(H)$ hysteresis loops, for different $H_{max}$ values, as indicated by the labels in panel A. Left and right columns correspond to $f=205$ and $765$ kHz, respectively. Each pair of colour panels corresponds to a different $K$ value (indicated within the figure) and its corresponding $d_{aggl}$ (table \ref{table-d_aggl}).}
\label{fig:MHcurves}
\end{figure}

We will analyse now the hyperthermia properties of the obtained threshold sizes for the different $K$ values. Since the roles of surface coating and media viscosity are not very significant in relation to $d_{aggl}$, we have focused, for simplicity, on the $K-d_{aggl}$ pairs summarized on Table \ref{table-d_aggl}, which would set an ideal limit. Thus, we simulated the dynamic hysteresis loops for the three cases considered, to then evaluate the heating capability. Some representative hysteresis loops are shown in Figure \ref{fig:MHcurves}.

\begin{figure}[h!]
\centering
\includegraphics[width = 1.0\columnwidth]{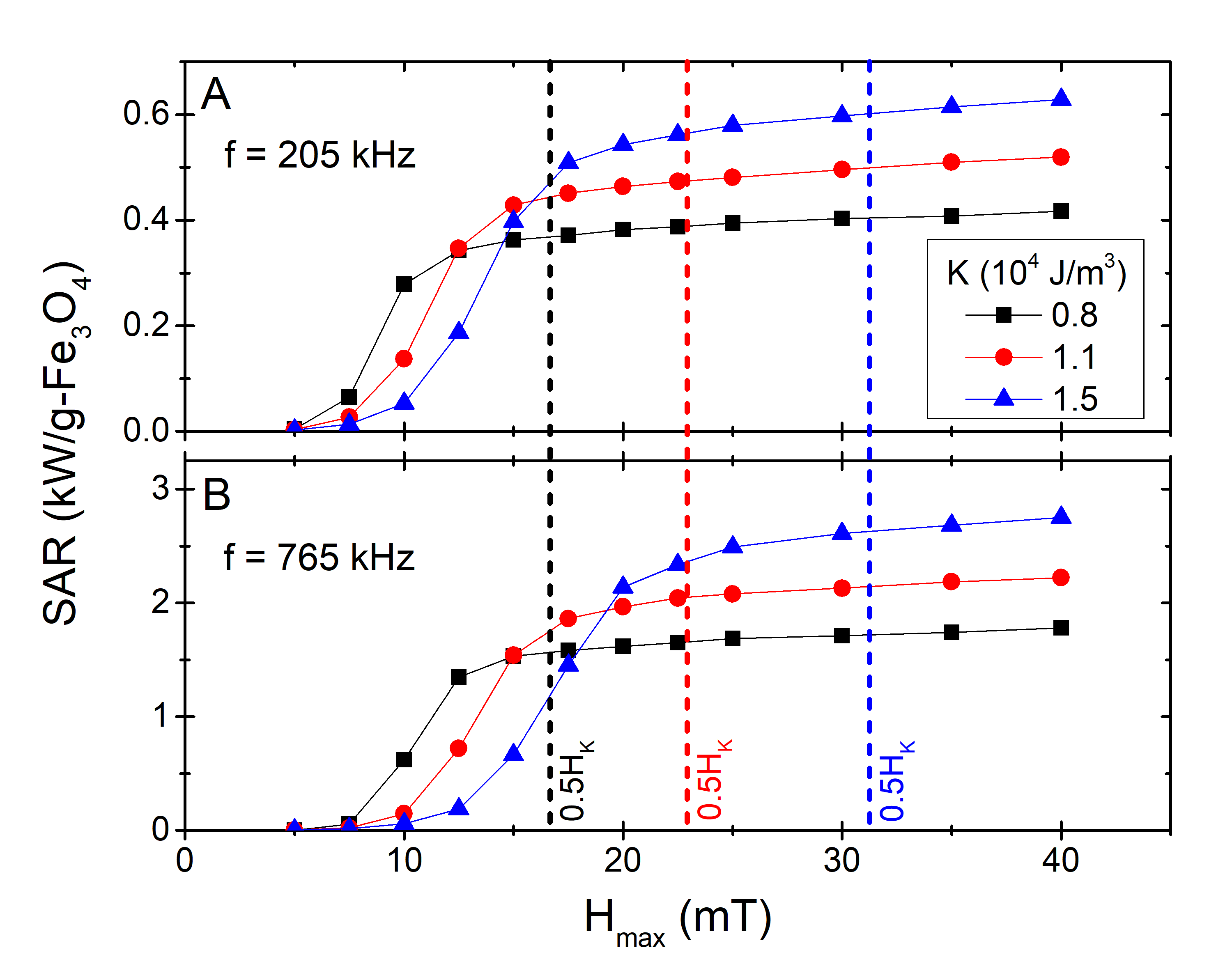}
\caption{SAR \textit{vs.} $H_{max}$ for the three $K$ values (at corresponding $d_{aggl}$), for $f=205$ and $765$ kHz. The vertical lines stand for half of the anisotropy field of each $K$ value (of same colour).}
\label{fig:SAR}
\end{figure}

The results displayed in Figure \ref{fig:MHcurves} show large differences depending on the value of $H_{max}$, illustrative of the minor-major loops competition \cite{Conde-Leboran2015,Munoz-Menendez2017}. This is further emphasized by the fact that higher frequency results in narrower loops for the small fields, but wider for the larger ones. The differences between the different $K$ cases are due to the different $H_{max}/H_{K}$ ratios, as discussed in Figure \ref{fig:MHcurves}. This is systematically analysed through the associated SAR values, shown in Figure \ref{fig:SAR}.
 
The results plotted in Figure \ref{fig:SAR} nicely fit within the general scenario discussed previously discussed (Figure \ref{fig:scaling}): larger $K$ allows higher SAR, provided enough field amplitude is reached (see corresponding $0.5H_{K}$ values -vertical dashed lines- for reference); for small $H_{max}$ values, however, it may occur that smaller-$K$ particles result in higher SAR due to the minor/major loops conditions, as discussed elsewhere \cite{Munoz-Menendez2017}. This is an important aspect to consider regarding the variation in \textit{local} heating due to size and/or anisotropy polydispersity \cite{Munoz-Menendez2015,Munoz-Menendez2017}), as the difference between \textit{blocked} and SPM particles would be the highest and thus also the \textit{locally} released heat \cite{Munoz-Menendez2017,Aquino2019}. The results are also clearly divergent from the \textit{linear response theory} model \cite{ROSENSWEIG2002}, for which $SAR\propto{H_{max}}^2$; this is not surprising as we are far from its applicability conditions (see e.g. Refs. \cite{Dennis2013,Carrey2011} for a detailed discussion).

\begin{figure}[h!]
\centering
\includegraphics[width = 1.0\columnwidth]{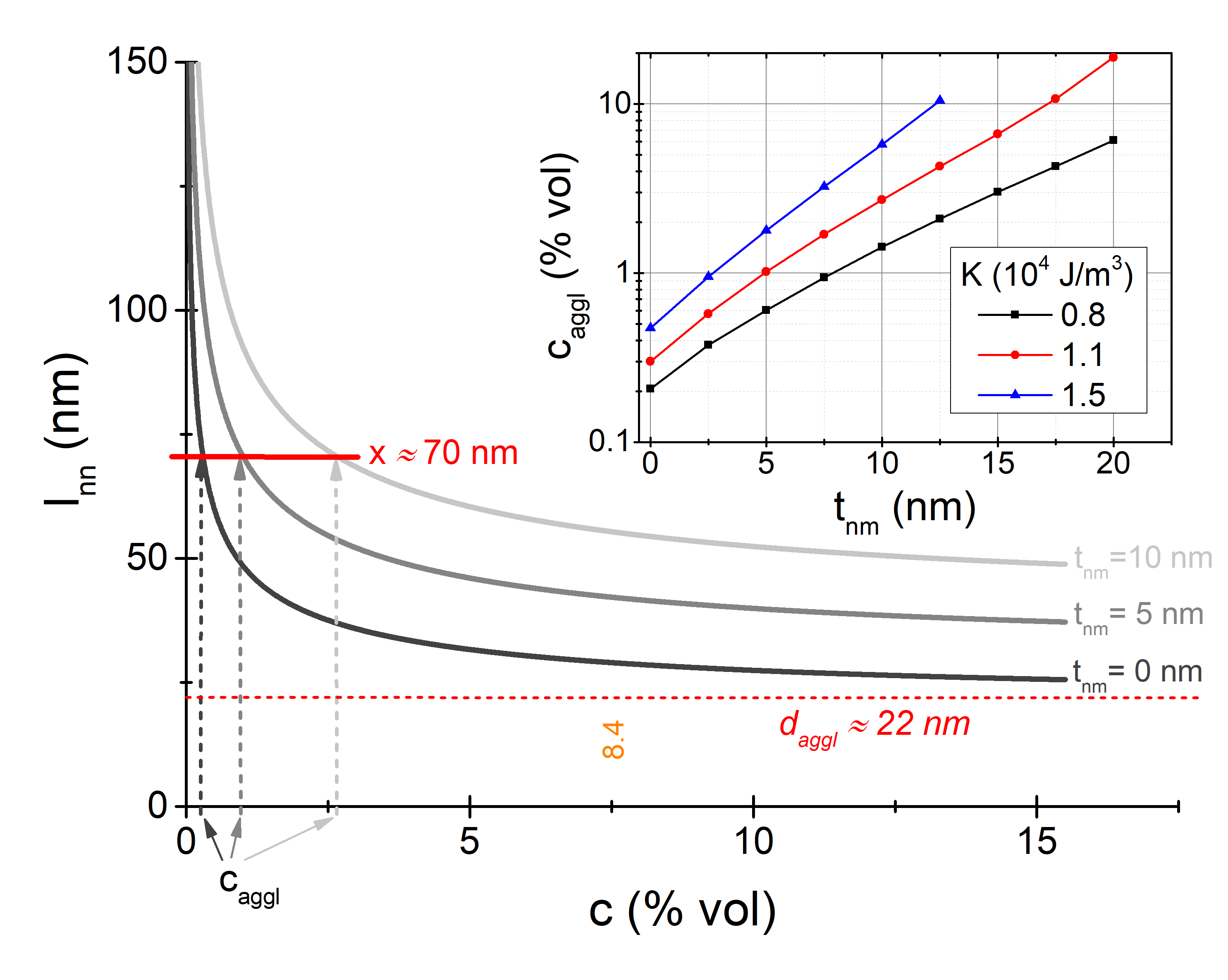}
\caption{$l_{NN}$ \textit{vs.} $c$ curves for different $t_{nm}$ values, for the $K=1.1*10^4$ $J/m^3$ case. The short (solid) horizontal line indicates the interparticle distance predicted by eq. \eqref{distance}, whereas the long (short-dashed) one indicates the $d_{aggl}$ value, to which $l_{NN}$ tends asymptotically. The vertical arrows indicate the corresponding $c_{aggl}$ values.  The inset shows $c_{aggl}$ \textit{vs.} $t_{nm}$ for the different values of $K$ considered.}
\label{fig:c_aggl}
\end{figure}

The predicted SAR values are quite large, implying that those particles would make efficient heat mediators. However, it is important to recall here that, so far, we made no considerations on the role of sample concentration. While this may appear reasonable as an initial approach, the fact is that the sample concentration is a key parameter to determine: first, because it defines the amount of deliverable heat; and second, because interparticle interactions (even without agglomeration) may significantly change the heating performance \cite{Serantes2010,Serantes2014,Branquinho2013,Conde-Leboran2015,Niculaes2017}.
To provide some hint on how the sample concentration, $c$ (\% volume fraction), relates to the assumptions made, we can consider it through the nearest-neighbors interparticle distance, $l_{NN}$. Following Tewari and Gokhale \cite{Tewari2004}, for a randomly distribution of monodisperse particles we can approximate $l_{NN}$ as \cite{Conde-Leboran2015}
\begin{equation}\label{lNN}
    l_{NN}=(d+2\cdot{t_{nm}})\cdot\frac{0.4465}{c^{1/3}}\left[{1+1.02625\left({\frac{c}{0.64}}\right)^{\frac{2}{3}}}\right].
\end{equation} Thus, by equating $l_{NN}$ to the diffusion distance $x$ (eq. \eqref{distance}) of the different $d_{aggl}$ values, we can obtain the related sample concentration threshold, $c_{aggl}$. This is shown in Figure \ref{fig:c_aggl}. 

The results shown in Figure \ref{fig:c_aggl} indicate that for bare particles ($t_{nm}=0$ nm) the applicability of the discussed arguments would be limited to very small concentrations, with $c_{aggl}\sim{0.2\%}$ for the $K=1.1*10^4$ $J/m^3$ case. However, the presence of a nonmagnetic coating significantly enlarges $c_{aggl}$, as illustrated in the main panel for the cases of $t_{nm}=5$ and 10 nm. This trend is systematically summarized within the inset, for the different values of $K$. It is observed that a coating of a few nanometers allows extending the applicability of our arguments within the ${1\%}-{10\%}$ range. It is interesting to notice how with higher $K$ this trend occurs with thinner $t_{nm}$, as expected due to the smaller $d_{aggl}$ sizes. At this point it is worth noting that for iron oxides it has been reported the existence of an essentially non-interacting regime at low concentrations \cite{SerantesPRB2010,Beola2020}, characteristic very attractive for the application viewpoint as it would allow discarding the complex role of interparticle interactions.

\section{Conclusions}\label{conclusions}

We have presented an estimation of the threshold sizes for magnetic agglomeration of magnetite-like nanoparticles, depending on their magnetic anisotropy. Our approach was based on the consideration that $K$ determines the stability of the particle magnetization and thus the likelihood of magnetic agglomeration, which involves physical translation and rotation of the particles themselves. By comparing the associated timescales, we have obtained that magnetite particles with usual anisotropy values should be relatively stable against agglomeration up to sizes in the range $\sim{20-25}$ nm in diameter. Then, we evaluated the associated hyperthermia performance, and found it to be relatively large (hundreds to thousands of W/g) for usual field/frequency conditions. The role of the nonmagnetic surface coating and that of the media viscosity appears secondary in determining the threshold sizes for agglomeration. 

The initial considerations were made with no considerations about sample concentration, despite being a critical parameter for the application. In this regard, simple estimates indicate that the assumptions would be strictly valid only for very diluted conditions. However, the presence of a nonmagnetic coating might significantly extent the validity of the approximations to higher concentrations (up to about $10\%$ volume fraction), showing that in this sense the nonmagnetic coating would play a key role.

It is important to recall that we have focused here on purely \textit{magnetic} agglomeration, i.e. an ideal assumption which does not consider the complex situation often found experimentally, where other forces -of electrostatic nature- often play a central role in the agglomeration \cite{Faraudo2013, BAKUZIS2013,Valleau1991} and lead to agglomeration at smaller sizes \cite{Gutierrez2019}. Including those falls however out of the scope of the present work, as it would result in a too complicated scenario. We have neither consider other important system characteristics as polydispersity in size (both regarding aggregation \cite{Balakrishnan2020} and heating \cite{Munoz-Menendez2015}), and in anisotropy. The latter is expected to play a key role based on its primary importance both for agglomeration and heating, as discussed here. However, to the best of our knowledge its role has only been investigated regarding heating performance \cite{Munoz-Menendez2017}, but not regarding agglomeration likelihood. Considering the combined influence of those parameters clearly constitutes a challenging task for future works.

Finally, it is necessary to recall the conceptual character of the present work: while we have considered magnetite-like values for $K$ and $M_S$ as a representative example, for simplicity those were taken as independent of size and temperature. However, it is well known that those may vary significantly within the size range of interest \cite{Demortiere2011}, and therefore the accurate determination of the agglomeration likelihood and hyperthermia performance would require including also those dependencies, together with the role of the nonmagnetic coating \cite{Goya2007}.

\section{Acknowledgements}
The authors acknowledge invaluable discussions and feedback from Prof. Roy Chantrell, Dr. Ondrej Hovorka, Dr. Luc\'ia Guti\'errez and Prof. Robert Ivkov. This work  used the computational facilities  at the Centro de Supercomputacion de Galicia (CESGA). D.S. acknowledges financial support from the Spanish \textit{Agencia Estatal de Investigaci\'on} (project PID2019-109514RJ-100). This research was partially supported by the Xunta de Galicia, Program for Development of a Strategic Grouping in Materials (AeMAT, Grant No. ED431E2018/08).

\bibliography{references.bib}

\begin{thebibliography}{53}
\bibitem [{are()}]{area8K}%
  \BibitemOpen
  \href@noop {} {}\bibinfo {note} {For a square loop, the coercive field is
  equal to the anisotropy field, $H_C=H_K$. Thus, since $H_K=2K/M_S$, the area
  is $A=(2H_K)*(2M_S)=8K$.}\BibitemShut {Stop}%
\end{thebibliography}
\end{document}